\documentclass[eqsecnum,showpacs,pre,twocolumn]{revtex4}
\usepackage{graphicx}
\usepackage{amsmath}
\usepackage{bm}
\usepackage{amsfonts}
\usepackage{amssymb}
\usepackage[usenames]{color}%
\begin{document}
\title{Attacks and Infections in Percolation Processes}
\author{Hans-Karl Janssen}
\affiliation{Institut f\"{u}r Theoretische Physik III,
Heinrich-Heine-Universit\"{a}t, 40225 D\"{u}sseldorf, Germany}
\author{Olaf Stenull}
\affiliation{Department of Physics and Astronomy, University of Pennsylvania,
Philadelphia, Pennsylvania 19104, USA}
\date{\today}

\begin{abstract}
We discuss attacks and infections at propagating fronts of percolation
processes based on the extended general epidemic process. The scaling behavior
of the number of the attacked and infected sites in the long time limit at the
ordinary and tricritical percolation transitions is governed by specific
composite operators of the field-theoretic representation of this process. We
calculate corresponding critical exponents for tricritical percolation in
mean-field theory and for ordinary percolation to 1-loop order. Our results
agree well with the available numerical data.
\end{abstract}
 \maketitle

\section{Introduction}

\label{sec:intro}

A very basic and largely open question in percolation theory is
following~\cite{Gra2012}: what are the scaling properties of the number of the
attacked and infected sites at propagating fronts of percolation processes in
the long time limit? Here, we study this question in the framework of dynamical
field theory of percolation.

To set the stage, to introduce concepts and define terms, it is helpful to
think initially in terms of the lattice model known as the extended general
epidemic process (EGEP)~\cite{JMS04}. Usually, dynamical percolation theory is
based on the general epidemic process (GEP)
\cite{Mol77,Bai75,Mur89,Gra83,Ja85,CaGra85,JaTa05}. Several years ago, we
generalized the GEP to the EGEP to allow, besides critical percolation, for
tricritical and first-order percolation transitions, which had been predicted
earlier from simulations in the context of depinning transitions of driven
surfaces in random media at zero temperature \cite{robbinsEtAl}.
Simulations also showed that tricritical and first-order percolation
disappear below three spatial dimensions in random media depinning
\cite{DroDa98} as well as in direct simulations of the EGEP
\cite{BiPaGra2012}.

\begin{figure}
\includegraphics[width=4.0cm]{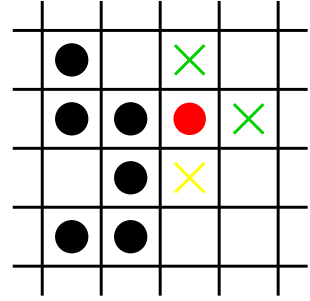}\caption{Part of a planar or spherical percolation
front propagating from left to right. Red: an activated cell (the
agent), black: inactive agents, green: susceptible sites, yellow: weakened
site. Both, the susceptible sites and the weakened site, are attackable. In the
next time step the agent is deactivated, and zero, one, two or three of the
attackable places become agents. Note the asymmetry of the number of the
attackable sites in the direction of the propagating front.}%
\label{Front}%
\end{figure}

The EGEP can be viewed as an extension of the well known
susceptible-infected-removed (SIR) model~\cite{kerMcK1927} and is also referred
to as the susceptible-weak-infected-removed (SWIR) model. It is given by the
reaction scheme
\begin{subequations}
\label{EGEP}%
\begin{align}
S(\mathbf{n})+I(\mathbf{m})\quad &  \overset{\alpha}{\longrightarrow}\quad
I(\mathbf{n})+I(\mathbf{m})\,,\label{reactA}\\
S(\mathbf{n})+I(\mathbf{m})\quad &  \overset{\beta}{\longrightarrow}\quad
W(\mathbf{n})+I(\mathbf{m})\,,\label{reactB'}\\
W(\mathbf{n})+I(\mathbf{m})\quad &  \overset{\gamma}{\longrightarrow}\quad
I(\mathbf{n})+I(\mathbf{m})\,,\label{reactC}\\
I(\mathbf{n})\quad &  \overset{\lambda}{\longrightarrow}\quad R(\mathbf{n})\,.
\label{reactD}%
\end{align}
\end{subequations}
with reaction rates $\alpha$, $\beta$, $\gamma$, and $\lambda$. $S$, $W$, $I$,
and $R$ respectively denote susceptible, weak, ill (activated), and
removed (dead, immune, the debris) individuals on nearest neighbor sites
$\mathbf{n}$ and $\mathbf{m}$. A susceptible individual may be infected by an
ill neighbor (agent) with rate $\alpha$ [reaction~(\ref{reactA})], or it may be
weakened without becoming ill with rate $\beta$ [reaction~(\ref{reactB'})]. 
An attack is defined as any such attempt to either infect or weaken a susceptible neighbor.
A further contact of a weak individual with an agent leads then to an
infection with rate $\gamma>\alpha$ [reaction~(\ref{reactC})]. By these
contagions, the disease spreads diffusively~\cite{footnote1}. Agents die with a
rate $\lambda$ [reaction~(\ref{reactD})]. Figure~\ref{Front} shows a sketch of
the process.
As mentioned above, the EGEP features first order, second order and tricritical phase transitions depending on the reaction rates. The second order transition belongs to the universality class of dynamic isotropic percolation (dIP), and the tricitical transition belongs to the universality class of tricritical dynamic isotropic percolation (TdIP). Through previous field theoretic studies~\cite{Ja85,CaGra85,JMS04}, much is known about the critical behavior of these universality classes. Most notably, the basic critical exponents such as exponents of order parameters, correlation length and the dynamical exponent have been calculated to 2-loop order.

In the following, we use the physical picture emanating from the EGEP to
identify operators in the field theory of dIP and TdIP that describe the scaling
properties of the number of the attacked and infected sites in the long time
limit. It turns out that these operators are composite operators (local products of fields and their derivatives) whose scaling behavior is {\em not} described by the known basic critical exponents of dIP and TdIP. We calculate the scaling exponents of these composite operators for the tricritical percolation transition in mean-field (MF) theory and for the
ordinary percolation transition to 1-loop order and we compare our results with
the available numerical data.

\section{Dynamical percolation and the extended general epidemic process}

Here, we briefly review some of the basics of dIP, TdIP and the EGEP
to provide some more background information, in particular, on their
interrelation.

\subsection{The essence of isotropic percolation}
\label{subsec:essence} The essence of isotropic percolation processes can be
summarized by four statements describing the universal features of the
evolution of such processes on a homogeneous substrate. Let us use the language
of an epidemic. Denoting the density of the agents (the activated substrate)
$I$ by $n(\mathbf{r},t)$ and the density of the debris $R$ by
$m(\mathbf{r},t)$, these four statements read:
\begin{enumerate}
\item[(i)] There is a manifold of absorbing states with $n\equiv0$ and
corresponding distributions of $m$ depending on the history of $n$. All these
absorbing states correspond to the extinction of the epidemic.
\item[(ii)] The substrate becomes activated (infected) depending on the
density of agents \emph{and} the density of the debris. This mechanism
introduces memory into the process. The debris ultimately stops the disease
locally. However, it is possible that the activation is strengthened by the
debris through some mechanism (e.g.~prior sensibilization of the substrate by
an exposure to the agents).
\item[(iii)] The process (the disease) spreads out diffusively by
contamination. The agents become deactivated to immune debris after a short
time.
\item[(iv)] There are no other slow variables. Microscopic degrees of freedom
can be summarized into a local noise or Langevin force $\zeta(\mathbf{r},t)$
respecting the first statement (i.e., the noise cannot generate agents).
\end{enumerate}

The general form of a Langevin equation formulating these statements is
\begin{subequations}
\label{eq:General Langevin}%
\begin{align}
\lambda^{-1}\dot{n}  &  =\nabla^{2}n+R(n,m)\,n+\zeta
\,,\label{eq:General Langevin-1}\\
m(\mathbf{r},t)  &  =\lambda\int_{-\infty}^{t}n(\mathbf{r},t^{\prime
})\,dt^{\prime}\,,
\end{align}
\end{subequations}
where $\lambda$ is a kinetic coefficient and the Gaussian noise correlation
reads
\begin{equation}
\overline{\zeta(\mathbf{r},t)\zeta(\mathbf{r}^{\prime},t^{\prime})}
=\lambda^{-1}Q\big(n,m\big)\,\delta(\mathbf{r}-\mathbf{r}
^{\prime})\,\delta(t-t^{\prime})\,.\label{LKraft}
\end{equation}

The dependence of the rate $R(n,m)$ on the density of the debris
$m(\mathbf{r},t)$ describes memory of the process mentioned above. We are
interested primarily in the behavior of the process close to percolation, where
$n$ and $m$ are small allowing for polynomial expansions $R(n,m)=-(\tau
+g_{1}m+g_{2}m^{2}+\ldots)n$, $Q(n,m)=(g_{3}+\ldots)n$. We will revisit this
Langevin equation in Sec.~\ref{subsec:responseFunctional}.

\subsection{The EGEP}

When the EGEP takes place on a finite lattice, the manifold of states
without any agent is inevitably absorbing. Whether a single initial agent leads
to an everlasting epidemic in an infinite system depends on the ratio
$\alpha/\lambda$. With $\lambda$ fixed and $\beta=0$, there is a certain value
$\alpha=\alpha_{c}$ such that for all $\alpha>\alpha_{c}$ an eternal epidemic
(a pandemic) occurs. The probability $P(\alpha)$ for the occurrence of a
pandemic as a function of $\alpha$ goes to zero continuously at the critical
point $\alpha_{c}$. The behavior of the process near this critical point is in
the universality class of dIP.

As we have shown~\cite{JMS04}, the occurrence of the weak individuals gives
rise to an instability that can lead to a discontinuous transition and compact
 growth of the epidemic if $\gamma$ is greater than a
critical value $\gamma_{c}(\alpha,\beta)$. In the enlarged three-dimensional
phase space spanned by $\alpha$, $\beta$, and $\gamma$ with fixed $\delta$,
there exists a critical surface associated with the usual continuous
percolation transition and a surface of first order transitions characterized
by a finite jump in the probability $P(\alpha,\beta,\gamma)$ for the occurrence
of a pandemic. These two surfaces of phase transitions meet at a line of
tricritical points determined in MF theory by
\begin{equation}
\gamma=\gamma_{c}=\alpha(1+\alpha/\beta)\,,\quad\alpha=\alpha_{c}=1/z\,,
\end{equation}
where $z$ is the coordination number of the lattice. The behavior of the
process near these tricritical points is in the universality class of
TdIP.  Rather precise numerical verifications of some of the
predictions of Ref.~\cite{JMS04} have been made by Bizhani, Paczuski and
Grassberger~\cite{BiPaGra2012}. These authors also showed that first
order and tricritical percolation disappears below three dimensions.

A MF description of the EGEP can be derived by treating the reaction equations
(\ref{EGEP}) as deterministic equations without fluctuations.  The latter is
described by the system of differential equations
\begin{subequations}
\label{meanfield}%
\begin{align}
\dot{S}(\mathbf{n},t)  &  =-(\alpha+\beta)S(\mathbf{n},t)\Sigma I(\mathbf{n}%
,t)\,,\label{meanS}\\
\dot{W}(\mathbf{n},t)  &  =\bigl[\beta S(\mathbf{n},t)-\gamma W(\mathbf{n}%
,t)\bigr]\Sigma I(\mathbf{n},t)\,,\label{meanW}\\
\dot{I}(\mathbf{n},t)  &  =\bigl[\alpha S(\mathbf{n},t)+\gamma W(\mathbf{n}%
,t)\bigr]\Sigma I(\mathbf{n},t)
\nonumber\\
&-\lambda I(\mathbf{n},t)\,,\label{meanI}\\
\dot{R}(\mathbf{n},t)  &  =\lambda I(\mathbf{n},t) \label{meanR}%
\end{align}
governing the dynamics of the different kinds of individuals. Here, $\Sigma X(\mathbf{n},t):=\sum_{\mathbf{m}}^{nn(\mathbf{n}%
)}X(\mathbf{m},t)$ denotes the summation over the nearest neighbors of
$\mathbf{n}$ of a quantity $X$ defined on lattice points. At each lattice site
there is the additional constraint
\end{subequations}
\begin{equation}
S(\mathbf{n},t)+W(\mathbf{n},t)+I(\mathbf{n},t)+R(\mathbf{n},t)=1\,,
\label{constraint}%
\end{equation}
if initially $S(\mathbf{n},0)=1$, $W(\mathbf{n},0)=I(\mathbf{n}%
,0)=R(\mathbf{n},0)=0$. Thus, $S(\mathbf{n},t)$, $W(\mathbf{n},t)$,
$I(\mathbf{n},t)$, and $R(\mathbf{n},t)$ can be interpreted as the
probabilities of finding the corresponding state at a site $\mathbf{n}$ at time
$t$. Note that this constraint is a defining feature of the EGEP. It
is valid beyond MF theory.

Equations~(\ref{meanS}) and (\ref{meanW}) are readily integrated as functions
of $\Sigma R(\mathbf{n},t)=\sum_{\mathbf{m}}^{nn(\mathbf{n})}R(\mathbf{m},t)$.
We obtain
\begin{equation}
S(\mathbf{n},t)=\exp\bigl[-(\alpha+\beta)\Sigma R(\mathbf{n}%
,t)\bigr] \label{sol1}%
\end{equation}
and
\begin{align}
W(\mathbf{n},t)&=\frac{\beta}{\alpha+\beta-\gamma}\Big\{\exp\bigl[-\gamma\Sigma
R(\mathbf{n},t)\bigr]
\nonumber\\
&-\exp\bigl[-(\alpha+\beta)\Sigma R(\mathbf{n}%
,t)\bigr]\Big\}\,, \label{sol2}%
\end{align}
where the time scale $\lambda$ has been set to unity for simplicity. In a
continuum approximation of the lattice points,
$\mathbf{n}\rightarrow\mathbf{x}$, the occupation numbers $S$, $W$, $I$, and
$R$ change into corresponding densities.
We set $R(\mathbf{n},t)\rightarrow m(\mathbf{x},t)$, $I(\mathbf{n}%
,t)\rightarrow n(\mathbf{x},t)\sim\dot{m}(\mathbf{x},t)$, $\Sigma
I(\mathbf{n},t)\rightarrow zn(\mathbf{x},t)+l^{2}\nabla^{2}n(\mathbf{x},t)$
with  $l$  being a length proportional to the lattice constant.
Equation~(\ref{meanI}) together with the solutions~(\ref{sol1}) and
(\ref{sol2}) then produces the mean field equation of motion of the EGEP ,
\begin{align}
\frac{\partial n(\mathbf{x},t)}{\partial t}&=\Bigg\{    \frac{(\alpha
+\beta)(\alpha-\gamma)}{\alpha+\beta-\gamma}\exp\bigl[-(\alpha+\beta
)zm(\mathbf{x},t)\bigr]\nonumber\\
&  +\frac{\beta\gamma}{\alpha+\beta-\gamma}\,\exp\bigl[-\gamma zm(\mathbf{x}%
,t)\bigr]\Bigg\}zn(\mathbf{x},t)
\nonumber\\
&-n(\mathbf{x},t)+\alpha l^{2}\nabla
^{2}n(\mathbf{x},t)\nonumber\\
&=-\Big\{    (1-z\alpha)+z^{2}\bigl[\alpha(\alpha+\beta)-\beta\gamma
\bigr]m(\mathbf{x},t)
\nonumber\\
&+O(m(\mathbf{x},t)^{2})\Big\}n(\mathbf{x},t)+\alpha
l^{2}\nabla^{2}n(\mathbf{x},t)\,. \label{mfEGEP}%
\end{align}
These MF equations should be compared with the deterministic part of the
Langevin-equation (\ref{eq:General Langevin}). In MF approximation, $\tau$ is
proportional to $(1-z\alpha)$, and $g_{1}$ is proportional to
$\bigl(\alpha(\alpha+\beta)-\beta\gamma\bigr)$ and $g_{2}$ and $g_{3}$ are
finite positive quantities. Hence $\tau$ vanishes at the percolation threshold,
and in addition $g_{1}$ is zero at the tricritical point.

\subsection{Response functional}
\label{subsec:responseFunctional}

Now, we return to the Langevin-equation (\ref{eq:General Langevin}).
Reformulating it as a dynamic response functional \cite{Ja76,DeDo76,Ja92,Uwe14}
leads to~\cite{JMS04}
\begin{align}
\mathcal{J}=\int d^{d}x\int dt\,\lambda\tilde{n}\big( & \lambda^{-1}\partial
_{t}-\nabla^{2}+\tau+g_{1}m
\nonumber\\
&+g_{2}m^{2}-g_{3}\tilde{n}\big)n\,, \label{J-EGEP}%
\end{align}
where $\tilde{n}(\mathbf{x},t)$ is the response field. This response functional
as it stands is strictly speaking not a minimal field theoretical model at this
stage as we have kept it general enough to encompass both the dIP and the TdIP
universality classes. We will review the parameter settings and rescalings that
take us to either universality class specifically in a moment. For the ease of
the argument, we will refer to $\mathcal{J}$ as the EGEP response functional.

Of course, this dynamic functional can be also derived by pursuing other
approaches. For example, one could proceed from the MF equations of motion of
the EGEP (\ref{mfEGEP}) and incorporate fluctuations by adding an absorbing
noise source $\zeta (\mathbf{x},t)$. This again leads to the full set of
equations (\ref{eq:General Langevin},\ref{LKraft}), and finally to the EGEP
response functional~(\ref{J-EGEP}). Or, as some readers might prefer, one can
recast the master-equation corresponding to the reaction scheme of the SWIR
(\ref{EGEP}) as a coherent state path integral (CSPI)-action
\cite{Doi76,GraSche80,Pel84,Ca97,Ca08,Uwe14,TaHoVo05,Wie15} and then integrate
out the coherent fields corresponding to $S$, $W$, and $R$ followed by a
na\"{\i}ve continuum approximation. After switching to number densities as
field variables via the Grassberger transformation and deletion of irrelevant
operators,  see e.g. \cite{JaTa05}, one also arrives at the functional
(\ref{J-EGEP}). However, we prefer the more universal approach outlined in
Sec.~\ref{subsec:essence} that boils down to a purely mesoscopic stochastic
formulation based on the correct order parameters identified through physical
insight in the nature of the critical phenomenon \cite{JMS04}.

The EGEP response functional contains a redundant parameter. This redundancy is
connected to the rescaling transformation
\begin{subequations}
\label{rescaling}%
\begin{align}
n  &  \rightarrow b\,n\,,\quad\tilde{n}\rightarrow b^{-1}\tilde{n}%
\,,\\
g_{1}  &  \rightarrow b^{-1}g_{1}\,,\quad g_{2}\rightarrow b^{-2}g_{2}\,,\quad
g_{3}\rightarrow bg_{3}\,, \label{Resc}%
\end{align}
\end{subequations}
($b$ is some scaling parameter,) that leaves $\mathcal{J}$ invariant. Note that
the combinations $g_{1}g_{3}$ and $g_{2}g_{3}^{2}$ are invariant under this
transformation.

At the tricritical point, where $g_{1}$ vanishes in MF theory, we fix the
redundancy by choosing $b=g_{3}^{-1}$, and for convenience we abbreviate $g =
g_{2}g_{3}^{2}$. There are two critical parameters, namely $\tau$ and
$\sigma=g_{1}g_{3}$. The fields scale as $\tilde{n}\sim \mu^{2}$,
$n\sim\mu^{d-2}$, and $m\sim\mu^{d-4}$ where $\mu$ is the inverse scaling
length and $d$ is the spatial dimension. The scaling of the coupling constant
$g\sim\mu^{2(5-d)}$ shows that the upper critical dimension is $d_{c}=5$.

Away from the tricritical point, $g_{1}$ is a finite quantity and $g_{2}$
becomes irrelevant. We chose $b=\sqrt{g_{1}/g_{3}}$ and set
$\sqrt{g_{1}g_{3}}=g$. The stochastic response functional is invariant under
the reflection transformation
$m(\mathbf{x},t)\leftrightarrow-\tilde{n}(\mathbf{x},-t)$. The  na\"{\i}ve
scaling is given by $\tilde{n}\sim m\sim\mu^{(d-2)/2}$, $n\sim\mu^{(d+2)/2}$,
and $g\sim\mu^{(6-d)/2}$ with $d_{c}=6$ as the upper critical dimension of
percolation.

For the MF phase diagram of the EGEP in terms of universal quantities, see
Fig.~1 of Ref.~\cite{JMS04}. Our critical parameter $\sigma$ here corresponds
to $-f$ there.

Further details on the field theory of the EGEP at the tricritical and the
percolation point can be found in Refs.~\cite{Ja85,JaTa05,JMS04}. For
completeness and for later use, we close our brief review of the EGEP by
restating the renormalization scheme that we have applied in the past:
\begin{subequations}
\begin{align}
n  &  \rightarrow\mathring{n}=Z^{1/2}n\,,\quad\mathring{\tilde{n}}%
\rightarrow\tilde{n}=\tilde{Z}^{1/2}\tilde{n}\,,
\\
 m&\rightarrow\mathring
{m}=\tilde{Z}^{1/2}m\,, \quad \lambda
\rightarrow\mathring{\lambda}=(\tilde{Z}/Z)^{1/2}\lambda\,,
\\
\tau&\rightarrow\mathring{\tau}=\tilde{Z}^{-1}Z_{\tau}\tau
+\mathring{\tau}_{c}\,,\quad g^{2}\rightarrow\mathring{g}^{2}=G_{\varepsilon
}^{-1}\tilde{Z}^{-3}Z_{u}u\mu^{6-d}\,, \label{RenSch}%
\end{align}
\end{subequations}
where $G_{\varepsilon}=\Gamma(1+\varepsilon)/(4\pi)^{d/2}$ and the open circles
indicate unrenormalized quantities.

\section{Attacks and Infections: Observables}

How to measure attacks and infections near the spreading percolation front? Let
us go back to the lattice formulation of the EGEP.  At a lattice point
$\mathbf{n}$ occupied by an agent, the probability that there are attackable
(susceptible or weak) individuals at a neighboring site is
\begin{equation}
S+W=1-z\alpha R+\cdots\,, \label{attprob}%
\end{equation}
where we have retained only the leading terms. Accounting for the probability
$I(\mathbf{n})$ that site $\mathbf{n}$ is actually occupied by an agent, the
probability of attacks in or against the direction $\boldsymbol{\delta}$ of the
propagating front is
\begin{equation}
A_{\pm}(\mathbf{n})=I(\mathbf{n})\bigl[1-z\alpha R(\mathbf{n\pm}%
\boldsymbol{\delta})\bigr]\,,
\end{equation}
The total number of attacks is therefore proportional to leading order to the
probability of agents $I(\mathbf{n})$ as one would expect since the agents have
fractal dimensions. The main quantity of interest is the difference between
attacks in and against the direction of spreading,
\begin{align}
A_{+}(\mathbf{n})-A_{-}(\mathbf{n})&=z\alpha I(\mathbf{n})\bigl[R(\mathbf{n-}%
\boldsymbol{\delta})-R(\mathbf{n+}\boldsymbol{\delta})\bigr]
\nonumber\\
&\approx-z\alpha
I(\mathbf{n})\boldsymbol{\delta}\cdot\mathbf{\nabla}R(\mathbf{n}) \, .
\end{align}
In the continuum approximation, the corresponding mean value of the
difference-number of attacks per agent at the spreading front is
therefore proportional to the attack-ratio%
\begin{equation}
\mathcal{A}(\mathbf{x},t)=\boldsymbol{\delta}\cdot\frac{\langle n(\mathbf{x}%
,t)\nabla m(\mathbf{x},t)\rangle}{\langle n(\mathbf{x},t)\rangle} \, .
\label{attrat}%
\end{equation}

What is the corresponding infection-ratio $\mathcal{I}(\mathbf{x},t)$? The rate
of infections of a neighbor of an agent at lattice point $\mathbf{n}$ is
given by the combination%
\begin{equation}
\alpha S+\gamma W=\alpha-\bigl[\alpha(\alpha+\beta)-\beta\gamma\bigr]R+O(R^{2}%
)\,. \label{infprob}%
\end{equation}
At the threshold of ordinary percolation, the coefficient $\sigma=\bigl(\alpha
(\alpha+\beta)-\beta\gamma\bigr)$ is a finite positive quantity, and the
infection probability (\ref{infprob}) is proportional to the probability of
attacks (\ref{attprob}) to leading order. Thus we find%
\begin{equation}
\mathcal{I}_{perc}(\mathbf{x},t)\sim
\mathcal{A}_{perc}(\mathbf{x},t)\,.
\label{percIA}%
\end{equation}
At the tricritical point, however, $\sigma$ is zero. Thus the infection
probability (\ref{infprob}) is determined by the second order term $\sim
R^{2}$. We obtain therefore%
\begin{equation}
\mathcal{I}_{tric}(\mathbf{x},t)=\boldsymbol{\delta}\cdot\frac{\langle
n(\mathbf{x},t)\nabla m(\mathbf{x},t)^{2}\rangle}{\langle n(\mathbf{x}%
,t)\rangle}\,, \label{infrat}%
\end{equation}
and a different infection-ratio%
\begin{equation}
\mathcal{I}_{tric}(\mathbf{x},t)\not\sim\mathcal{A}_{tric}(\mathbf{x},t)\,.
\label{tricIA}%
\end{equation}

\section{Attacks and Infections: Scaling Behavior}

Our discussion in the previous section implies that the scaling behavior of attacks and infections is determined by the operator products $n(\mathbf{x},t)\nabla m(\mathbf{x}%
,t)$ and $n(\mathbf{x},t)\nabla m(\mathbf{x},t)^{2}$. Here, we will study their
behavior under the renormalization group.

As a prelude, we first extract the MF scaling behavior. At the ordinary
percolation transition, na\"{\i}ve scaling gives $n\nabla m\sim\mu^{d+1}$ which
together with  $n\sim\mu^{d/2+1}$ leads to
$\mathcal{I}_{perc}=\mathcal{A}_{perc}\sim\mu^{d/2}$.  At the percolation
front, we have $\mathbf{x}\sim t^{z}$ if the percolation starts near
$\mathbf{x}=0$ at time $t=0$. With $\mathbf{x}\sim\mu^{-1}$ and the na\"{\i}ve
$z=2$, we therefore obtain
 $\mathcal{I}_{perc}=\mathcal{A}_{perc}\sim t^{-d/4}$. Hence the
MF scaling behavior at and above the upper critical dimension
$d_{c}=6$ is%
\begin{equation}
\mathcal{I}_{perc}^{(mf)} \sim \mathcal{A}_{perc}^{(mf)}\sim t^{-3/2}%
\end{equation}
in full agreement with Grassbergers simulations~\cite{Gra2012}. At the
tricritical point, the same type of reasoning leads to
\begin{equation}
\mathcal{I}_{tric}^{(mf)}\sim t^{-3/2}\,,\quad\mathcal{A}_{tric}^{(mf)}\sim
t^{-1}%
\end{equation}
 at and above the upper critical dimension $d_{c}=5$.  Again the scaling exponents here are in full agreement with the available simulation results~\cite{Gra2012}.

Now, we determine the anomalous contributions to the scaling exponents arising
below the upper critical dimension. Here we will focus on the ordinary
percolation transition. At the tricritical point, one typically has to work to
two-loop order to get the leading anomalous contributions which is beyond the
scope of the present paper.

Under the renormalization process, the composite operator $n\nabla m$ induces
several other operators. In minimal renormalization (dimensional regularization
in conjunction with minimal subtraction), the induced operators that we have to
worry about are local products of $n$, $m$, $\tilde{n}$ and their derivatives
with equal or lower na\"{\i}ve dimension and same symmetries and conservation
properties as $n\nabla m$. Since the factor $n$ can only induce operators which
vanish with $n$, and the factor $\tilde{n}$ operators which vanish with
$\tilde{n}$, the complete list of
 independent composite operators that we have to include in our calculation reads
\begin{align}
(\mathcal{O}_{1},\ldots,\mathcal{O}_{7})=&(n\nabla m,n\nabla\tilde{n}%
,\nabla(nm),\nabla(n\tilde{n}),\nabla n,
\nonumber\\
&\nabla\nabla^{2}n,\nabla\partial _{t}n)\,. \label{opList}
\end{align}
In general, all the operators mix under the action of the renormalization group
and therefore it takes a matrix $Z_{\alpha\beta}$ of renormalization factors,
\begin{equation}
\lbrack\mathcal{O}]_{\alpha}=\sum_{\beta}Z_{\alpha\beta}\mathcal{O}_{\beta} \,
.
\label{RenOp}%
\end{equation}
to remove all superficial divergencies. To calculate this matrix, one can
augment the response functional with the additional interaction
\begin{align}
\delta\mathcal{J}&=\int d^{d}x\int dt\,\mathring{\lambda}\sum_{\alpha}%
\mathring{f}_{\alpha}\mathcal{\mathring{O}}_{\alpha}
\nonumber\\
&=\int d^{d}x\int
dt\,\lambda\sum_{\alpha,\beta}f_{\alpha}Z_{\alpha\beta}\mathcal{O}_{\beta }
\nonumber\\
&=\int d^{d}x\int dt\,\lambda\sum_{\alpha}f_{\alpha}[\mathcal{O}]_{\alpha}\,,
\end{align}
and then determine all the components of the renormalization matrix from the
relation between bare coupling constants $\mathring{f}_{\alpha}$ and their
renormalized counterparts $f_{\alpha}$.

Note, however, that the operators $m\nabla n$ and $\tilde{n}\nabla n$ are given
by combinations of the others. The total derivatives of $nm$ and $n\tilde{m}$,
as well as their renormalizations are related to the last two operators in the
list through insertions of the so-called equation of motions
$\delta\mathcal{J}/\delta\tilde{n}$ and $\delta\mathcal{J}/\delta n$. Of
course, the renormalizations of the bare $\mathcal{\mathring{O}}_{5}$ to
$\mathcal{\mathring{O}}_{7}$ are given by the renormalization constant $Z$.
Taken together, these observations imply that the matrix $Z_{\alpha\beta}$ has
trigonal structure, $Z_{\alpha\beta}=0$ if $\alpha>\beta$. Thus, to calculate
the scaling exponent of $\mathcal{O}_{1}=n\nabla m$, the only new [i.e., not
already contained in the renormalization scheme~(\ref{RenSch})] renormalization
factor that we have to extract is $Z_{11}$. In other words, although
$\mathcal{O}_{1}$ induces a bunch of other operators, these do not in turn have
an impact on its scaling behavior and can therefore be discarded (as long as we
focus on the scaling of $\mathcal{O}_{1}$ only). In a different context, we
have referred to operators of this kind as master operators~\cite{SteJa2000}.

For the actual calculation of $Z_{11}$, we proceed  as follows. Keeping only $f_{1}\neq0$, we find $Z_{11}$ from%
\begin{align}
\tilde{Z}\mathring{f}_{\alpha}  &  =f_{1}Z_{1\alpha}\,,\quad\alpha
=1,\ldots,4\,,\nonumber\\
\tilde{Z}^{1/2}\mathring{f}_{\alpha}  &  =f_{1}Z_{1\alpha}\,,\quad
\alpha=5,\ldots,7\,.
\end{align}
\begin{figure}
\includegraphics[width=8cm]{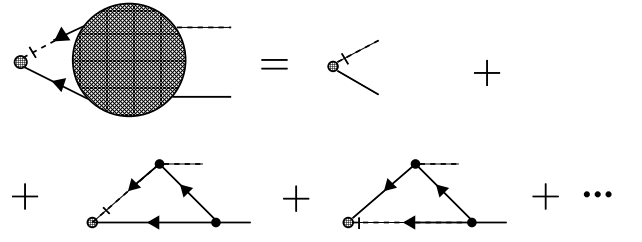}\caption{Diagrams needed for calculation the scaling exponent of  $\mathcal{O}_{1}$ to 1-loop order.}%
\label{1-Loop Dia}%
\end{figure}
Our 1-loop calculation, see the diagrams in Fig.~\ref{1-Loop Dia}, yields
\begin{equation}
\Gamma_{\mathcal{O}_{1}}=\lambda f_{1}\left(Z_{11}+\frac{u}{2\varepsilon}%
+\ldots \right)\mathcal{O}_{1}\,,
\end{equation}
where the ellipsis denote terms that are finite for
$\varepsilon=6-d\rightarrow0$. We obtain
\begin{equation}
Z_{11}=1-\frac{u}{2\varepsilon}+O(u^{2})\,.\label{Z11}%
\end{equation}
The renormalization of $\mathcal{O}_{1}$ in detail reads
\begin{equation}
\lbrack\mathcal{O}]_{1}=\sum_{\beta}Z_{1\beta}\mathcal{O}_{\beta}=(Z\tilde
{Z})^{-1/2}Z_{11}\mathcal{\mathring{O}}_{1}+\ldots\,,
\end{equation}
where the ellipsis here denotes the combination of the other
$\mathcal{O}_{\beta}$ that is not required for the calculation of the scaling
dimension of the attack-ratio. Applying the
renormalization-group differential operator%
\begin{equation}
\mathcal{D}_{\mu}\cdots:=\left.  \mu\frac{\partial\cdots}{\partial\mu
}\right\vert _{0}=\Bigg(\mu\frac{\partial}{\partial\mu}+\beta\frac{\partial
}{\partial u}+\zeta\lambda\frac{\partial}{\partial\lambda}+\kappa\tau
\frac{\partial}{\partial\tau}\Bigg)\cdots\,,
\end{equation}
to the attack-ratio (the RG-functions $\beta$, $\zeta$, and $\kappa$ can be
found in Refs.~\cite{Ja85,JaTa05}), we get
\begin{equation}
\bigl(\mathcal{D}_{\mu}-\omega_{1}\bigr)\mathcal{A}_{perc}(\mathbf{x}%
,t,\tau;u,\lambda,\mu)=0\,,\label{RGG-A}%
\end{equation}
with the exponent $\omega_{1}=\omega_{11}$ at the fixed point
$u=u_{\ast}=2\varepsilon/7+O(\varepsilon^{2})$ given by
\begin{equation}
\omega_{1}=\mathcal{D}_{\mu}\ln(Z_{11}/\tilde{Z}^{1/2})=\frac{7}{12}%
u+O(u^{2})=\frac{\varepsilon}{6}+O(\varepsilon^{2})\, .
\end{equation}
Dimensional analysis of $\mathcal{A}_{perc}$ yields%
\begin{equation}
\mathcal{A}_{perc}(\mathbf{x},t,\tau;u,\lambda,\mu)=\mu^{d/2}\mathcal{A}%
_{perc}(\mu\mathbf{x},\lambda\mu^{2}t,\mu^{-2}\tau;u,1,1)\,.
\end{equation}
Together with the solution of the RGG (\ref{RGG-A}) %
\begin{equation}
\mathcal{A}_{perc}(\mathbf{x},t,\tau;u,\lambda,\mu)=l^{\omega_{1}}%
\mathcal{A}_{perc}(\mathbf{x},t,l^{\kappa}\tau;u_{\ast},l^{\zeta}\lambda
,l\mu)\,,
\end{equation}
at the fixed point with $l$ being the dimensionless flow-parameter,
we obtain %
\begin{align}
&\mathcal{A}_{perc}(\mathbf{x},t,\tau;u,\lambda,\mu)
\nonumber\\
& =l^{\omega_{1}%
+d/2}\mathcal{A}_{perc}(l\mathbf{x},l^{z}t,l^{-1/\nu}\tau;u_{\ast},\lambda
,\mu)\nonumber\\
& =t^{-\alpha_{\mathcal{A}}}F(\mathbf{x}/t^{1/z},\tau t^{1/\nu z})\,,
\end{align}
where $F$ is some scaling function, for the scaling behavior of the
attack-ratio.
Its scaling exponent is%
\begin{equation}
\alpha_{\mathcal{A}}=\Bigg(\frac{d}{2}+\omega_{1}\Bigg)/z=\frac{3}%
{2}\Bigg[1-\frac{\varepsilon}{36}+O(\varepsilon^{2})\Bigg]\,,\label{scalexp}%
\end{equation}
where we used the expansion $z=2-\varepsilon/6+O(\varepsilon^{2})$.

Finally, we compare our result to the available numerical data. Currently, we know of such data only for $d=2$, where $\alpha_{\mathcal{A}}=0.518$~\cite{Gra2012}. Extrapolating our 1-loop result for
$\alpha_{\mathcal{A}}$ as it stands down to $d=2$ is problematic, of course. Thus, we resort to one of the established procedures for improving $\varepsilon$-expansion results, namely rational approximation. The general idea behind this technique is to augment an $\varepsilon$-expansion with higher order terms to incorporate additional information.
To this end, we note that in one-dimensional percolation, the attack-ratio is independent of time: microscopically it is simply $1$. Hence, we know rigorously that
$\alpha_{\mathcal{A}}=0$ in $d=1$. We incorporate this feature into our result by multiplying the right hand side of Eq.~(\ref{scalexp}) with the factor $\bigl[1-(\varepsilon/5)^{2}\bigr]$ resulting in the interpolation formula%
\begin{equation}
\alpha_{\mathcal{A}}=\frac{3(d-1)}{10}\Bigg(1+\frac{31}{180}\varepsilon
\Bigg)\,.\label{interpol}%
\end{equation}
Note that the extra factor was set up with $\varepsilon$ appearing quadratically and that therefore Eq.~(\ref{scalexp}) and (\ref{interpol}) are in absolute agreement to first order in $\varepsilon$ as they should. For $d=2$, our interpolation produces $\alpha_{\mathcal{A}}=0.507$ which is satisfyingly close to the numerical result stated above.

\section{Concluding Remarks}
In summary, we have discussed attacks and infections at percolating fronts based
on renormalized dynamical field theory of the EGEP. Important observables
measuring these attacks and infections are related to composite operators of
this theory that had not been studied hitherto. For the ordinary percolation
transition, we calculated the corresponding anomalous exponents to 1-loop
order. For the tricritical percolation transition, one has to go at minimum to
2-loop order to get anomalous contributions beyond MF theory. We leave this
interesting and challenging problem for future work.

\begin{acknowledgments}
This work was supported by the NSF under No.~DMR-1104701 (OS) and
No.~DMR-1120901 (OS).
\end{acknowledgments}


\begin{thebibliography}{99}
\bibitem{footnote1}
Note it is not the infected individuals but the disease as a whole that spreads
diffusively.
\end{thebibliography}
\end{document}